\documentclass[conference]{IEEEtran}
\IEEEoverridecommandlockouts
\usepackage{cite}
\usepackage{amsmath,amssymb,amsfonts}
\usepackage{graphicx}
\usepackage{textcomp}
\usepackage{hyperref}
\usepackage{xcolor}
\def\BibTeX{{\rm B\kern-.05em{\sc i\kern-.025em b}\kern-.08em
    T\kern-.1667em\lower.7ex\hbox{E}\kern-.125emX}}

\begin{document}

\title{When Machines Lie Differently: Detecting AI vs Human Fake News\\
{\footnotesize \textsuperscript{}}
}

\author{
\IEEEauthorblockN{Calvin Ibenye$^{1}$, Aya-Vera Jimenez$^{2,3}$, Samuel Jaeger$^{3}$, Dhrubajyoti Ghosh$^{3}$}
\IEEEauthorblockA{
$^{1}$Department of Computer Science, Kennesaw State University, Kennesaw, GA\\
$^{2}$Department of Mathematics, Kennesaw State University, Kennesaw, GA\\
$^{3}$School of Data Science and Analytics, Kennesaw State University, Kennesaw, GA}
}

\maketitle

\begin{abstract}
The rapid advancement of large language models has introduced a new class of AI-generated fake news that coexists with traditional human-written misinformation, raising fundamental questions about the detectability of different sources of deceptive content. In this study, we investigate whether the detectability of fake news depends on its origin by constructing two controlled binary classification tasks: distinguishing real news from human-written fake news and from AI-generated fake news. Using a feature representation that captures lexical diversity, readability, and emotional characteristics, and applying logistic regression, random forests, support vector machines, gradient boosting, neural networks, and an ensemble model, we evaluate performance using the area under the receiver operating characteristic curve (AUC). Across all models, we observe a consistent and substantial performance gap, with near-perfect AUC for AI-generated fake news and lower accuracy for human-written fake news. Feature-level analysis reveals that AI-generated fake news exhibits more uniform patterns in readability and emotional expression, reducing its overlap with real news. These findings highlight a critical asymmetry in misinformation detection and underscore the importance of accounting for the source of misinformation, and for the distinction between authorship and veracity, when designing detection systems.
\end{abstract}


\section{Introduction}
The rapid proliferation of misinformation has emerged as a critical challenge in the modern digital information ecosystem, with significant implications for public health, political processes, and societal trust. The widespread dissemination of false or misleading information has been shown to influence public opinion, amplify polarization, and undermine institutional credibility \cite{r1,r2,r3,r4}. Empirical evidence further suggests that false information spreads more rapidly and broadly than truthful content, largely due to its novelty and emotional appeal \cite{r4}. These findings have motivated a substantial body of research on automated fake news detection using techniques from machine learning and natural language processing.

Traditional approaches have primarily focused on distinguishing human-written fake news from legitimate news articles, relying on linguistic and stylistic features such as lexical diversity, syntactic structure, readability, and sentiment or emotional tone \cite{r5,r6}. Classical models such as logistic regression, support vector machines, and ensemble methods \cite{r21, r30} have been widely applied in this context, often achieving strong performance on benchmark datasets \cite{r7,r8}. A key underlying assumption is that fake news exhibits identifiable deviations from real news in writing style, emotional intensity, or structural coherence.

Recent advances in large language models (LLMs) \cite{r9} have introduced a second, distinct source of misinformation: AI-generated fake news. Modern generative systems can produce highly fluent, coherent, and contextually appropriate text at scale, closely mimicking human writing style, which raises questions about whether detection methods built around human-authored deception transfer to machine-authored deception. Studies on the detectability of AI-generated text have shown that detection models often generalize poorly across unseen generators or domains \cite{r10,r11}, but most existing work studies AI-generated versus human-written text in isolation \cite{r12,r13,r14, r29}, without directly comparing how \emph{detectable} human-written and AI-generated fake news are relative to one another under an identical modeling pipeline.

This distinction matters because the underlying generative mechanisms differ substantially. Human-written fake news is often crafted with explicit intent to persuade or mislead, leveraging emotional triggers such as fear, anger, or outrage \cite{r15}. AI-generated fake news, by contrast, arises from probabilistic language modeling, which may produce more neutral tone, smoother sentence structure, and reduced stylistic variability. If the two sources differ substantially in detectability, a single unified detection strategy may be insufficient, and adaptive or source-aware approaches may be required instead.


Related work in this space includes ensemble learning for distinguishing AI-generated from human-written fake news directly \cite{r12}, cross-prompt generalization of AI-fake-news detectors \cite{r30}, zero-shot detection based on log-probability curvature \cite{r31}, and adversarially trained detectors designed for robustness to paraphrasing \cite{r32}; these approaches primarily target authorship classification or generalization rather than veracity. Our study differs in framing: rather than classifying authorship, we hold the feature representation, preprocessing, and classifiers fixed and compare the \emph{separability of each source from real news}, isolating how the source of misinformation affects detectability rather than how to tell the two sources apart.

In this study, we conduct a controlled comparison of two detection tasks: (i) real vs.\ human-written fake news, and (ii) real vs.\ AI-generated fake news, using an identical feature set, preprocessing pipeline, and set of classifiers for both. 
This design holds the feature extraction, preprocessing, and classifiers fixed across the two tasks, enabling a direct comparison of their observed separability.

\section{Method}

Real news articles were sampled from the Politifact training dataset, while human-written fake-news articles were taken from its fake-news corpus. AI-generated fake-news articles were created by prompting ChatGPT to rewrite the original fake articles while preserving the same false claims and overall storyline, but changing the wording, sentence structure, and ordering. We then constructed two balanced binary classification datasets: $\mathcal{D}_H$, comparing real news with human-written fake news, and $\mathcal{D}_A$, comparing the same real-news sample with AI-generated fake news. Articles with missing or empty text were removed before feature extraction.

Each document is mapped to a feature vector $\phi(x)\in\mathbb{R}^p$ through a deterministic transformation designed to capture complementary aspects of textual structure and content. Lexical diversity is quantified using the type--token ratio \cite{r16},
\begin{equation}
\text{TTR} = \frac{|\text{unique tokens}|}{|\text{total tokens}|},
\end{equation}
which reflects vocabulary richness and serves as a proxy for stylistic diversity. Structural complexity is captured through standard readability indices, including the Flesch Reading Ease score \cite{r17},
\begin{equation}
\text{Flesch} = 206.835 - 1.015\left(\tfrac{\text{words}}{\text{sentences}}\right) - 84.6\left(\tfrac{\text{syllables}}{\text{words}}\right),
\end{equation}
together with the Flesch--Kincaid grade level \cite{r18}, SMOG \cite{r19}, and Coleman--Liau index \cite{r20}, which collectively encode structural regularity based on sentence length, word length, and character-level statistics. In addition, affective features are derived from the NRC emotion lexicon (anger, fear, joy, sadness, trust, anticipation, disgust, surprise) together with overall positive and negative sentiment, capturing the emotional intensity and distribution of affective cues linked to the propagation of misleading content. The final feature vector concatenates lexical, readability, and emotion-based features; observations with missing values are excluded.

Given this representation, we consider a family of supervised classifiers that estimate $P(y=1\mid\phi(x))$: logistic regression, support vector machines (with kernel transformations of the feature space), random forest and XGBoost (tree ensembles approximating nonlinear decision boundaries), a feed-forward neural network, and an ensemble predictor that averages the predicted probabilities of the five base models,
\begin{equation}
\hat{p}_{\text{ens}}(x) = \frac{1}{M}\sum_{m=1}^{M} \hat{p}_m(x).
\end{equation}
For each of $\mathcal{D}_H$ and $\mathcal{D}_A$, observations are randomly partitioned into training (80\%) and test (20\%) subsets; all preprocessing, including feature standardization, is fit on the training data only and then applied to the test data to avoid information leakage. Model performance is evaluated using AUC, the probability that a randomly chosen fake-news article receives a higher predicted score than a randomly chosen real-news article, which is invariant to classification thresholds and allows direct comparison across tasks. Because the feature extraction, preprocessing, and model classes are identical across $\mathcal{D}_H$ and $\mathcal{D}_A$, any systematic difference in AUC is attributable to the underlying text distributions rather than to modeling choices. Full dataset composition (sample sizes, class balance, topic and time coverage, article-length statistics) and the LLM generation configuration (model, prompt template, decoding parameters) are reported in the supplementary material.

\section{Results}
Table~\ref{tab:auc} reports test-set AUC for all six models on both tasks. The AI-generated fake news task is uniformly easier: AUC exceeds 0.99 for five of six models (ensemble: 0.9962), versus 0.89--0.96 for the human-written fake news task. A paired $t$-test confirms the gap is significant ($t=8.67$, $p<0.001$; mean improvement $\approx 0.049$), corroborated by a Wilcoxon signed-rank test ($p=0.016$).

\begin{table}[htbp]
\caption{Test AUC by model and task}
\begin{center}
\begin{tabular}{lcc}
\hline
\textbf{Model} & \textbf{Real vs.\ Human Fake} & \textbf{Real vs.\ AI Fake} \\
\hline
Logistic Regression & 0.9552 & 0.9964 \\
Random Forest        & 0.9475 & 0.9955 \\
SVM                   & 0.9520 & 0.9960 \\
XGBoost               & 0.9477 & 0.9961 \\
Neural Network        & 0.8878 & 0.9635 \\
Ensemble              & 0.9597 & 0.9962 \\
\hline
\end{tabular}
\label{tab:auc}
\end{center}
\end{table}

\begin{figure}[htbp]
\centerline{\includegraphics[width=0.95\linewidth]{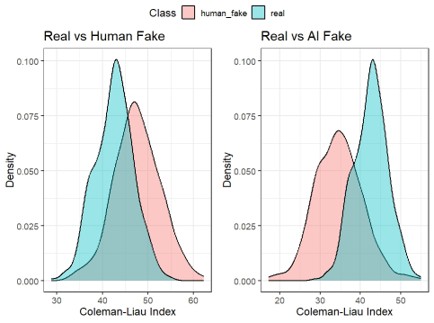}}
\caption{Distribution of the Coleman--Liau readability index for real and fake news. AI-generated fake news shows reduced overlap with real news relative to human-written fake news.}
\label{fig:density}
\end{figure}

Fig.~\ref{fig:density} shows the empirical density of the Coleman--Liau index for both tasks. Human-written fake news overlaps substantially with real news, whereas AI-generated fake news is more clearly separated, consistent with reduced stylistic variability in machine-generated text.

\begin{figure}[htbp]
\centerline{\includegraphics[width=0.95\linewidth]{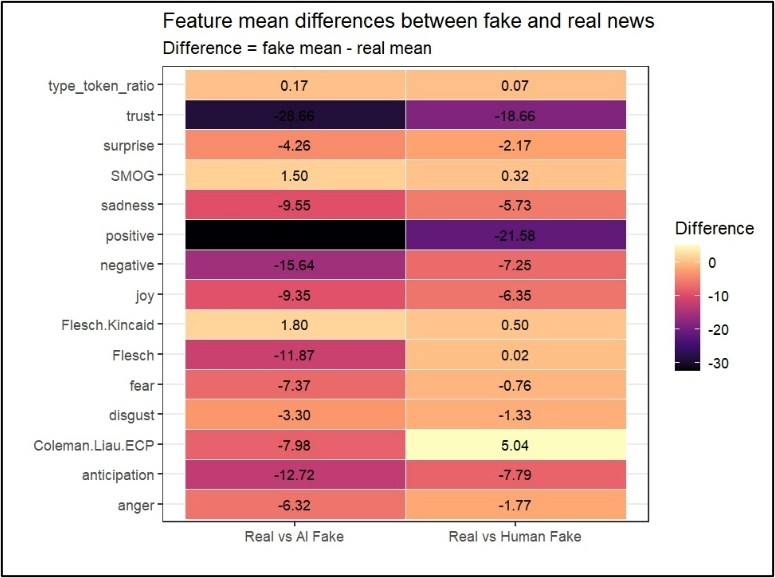}}
\caption{Feature-wise mean differences (fake $-$ real) by task. Deviations are larger for AI-generated fake news, indicating greater departure from real news.}
\label{fig:heatmap}
\end{figure}

Fig.~\ref{fig:heatmap} decomposes this separation by feature, plotting the mean difference between fake and real news for each task. Deviations are consistently larger in magnitude for AI-generated fake news, particularly for the type--token ratio and sentiment-related features (trust, positive sentiment), indicating a more uniform, less emotionally varied linguistic profile than human-written fake news.

\begin{figure}[htbp]
\centerline{\includegraphics[width=0.95\linewidth]{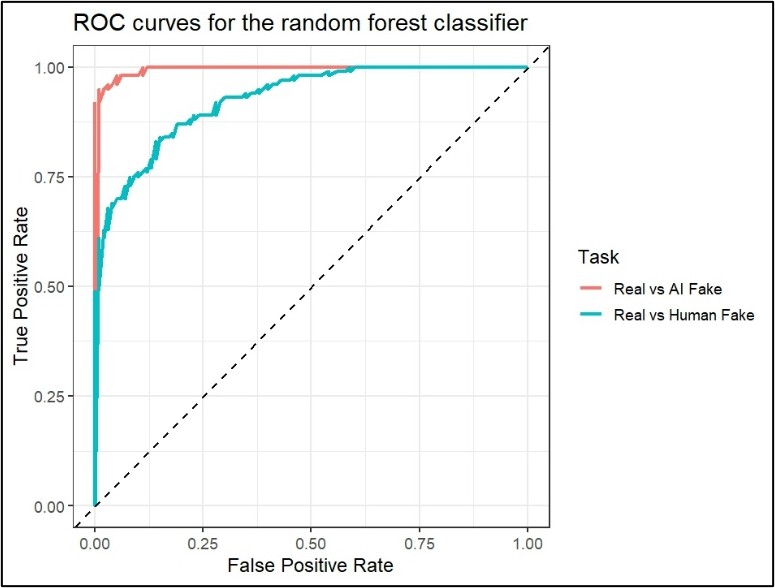}}
\caption{ROC curves for the random forest classifier. The real-vs-AI-fake curve dominates the real-vs-human-fake curve, indicating uniformly better detection of AI-generated fake news.}
\label{fig:roc}
\end{figure}

Fig.~\ref{fig:roc} shows ROC curves for the random forest classifier on both tasks; the real-vs-AI-fake curve dominates the real-vs-human-fake curve at every false positive rate, confirming the AUC gap holds across the full range of decision thresholds and not just at a single operating point.

\section{Discussion}
The consistency of the AUC gap across linear, kernel, tree-based, and ensemble models indicates that the difference is encoded in the feature space itself rather than tied to a specific decision boundary. This is consistent with prior findings in text classification and stylometry, where certain statistical features remain discriminative across a wide range of classifiers \cite{r25,r26}. The feature-level evidence (Fig.~\ref{fig:heatmap}) suggests AI-generated text currently follows more uniform patterns in sentence construction, lexical choice, and affective expression than either real news or human-crafted fake news, consistent with reports that machine-generated text lacks the ``burstiness'' of human writing \cite{r22,r23}. Human-written fake news, by contrast, is authored with intent to deceive and appears to more closely mimic the heterogeneity of legitimate journalism, likely because it incorporates stylistic cues that resemble legitimate reporting \cite{r24}, which plausibly explains its lower detectability. This asymmetry has practical implications: the high detectability of AI-generated fake news under relatively simple feature representations suggests current systems may be well-equipped to identify such content in its present form, but this advantage may be temporary, as advances in generative modeling are likely to reduce these detectable discrepancies over time \cite{r27}.

An important limitation of the present design is that the real-vs-AI-fake task confounds \emph{veracity} with \emph{authorship}: real articles are human-written and fake articles are AI-generated, so part of the observed separability could reflect generic AI-authorship artifacts (e.g., decoding regularities, prompt formatting, length) rather than fakeness per se. We do not rule this out here. Disentangling the two would require a four-way design (human-real, human-fake, AI-real, AI-fake) that classifies authorship and veracity separately; we treat this as the primary direction for follow-up work rather than a claim already established in this short paper, and we report additional surface-form and metadata baselines toward this question in the supplementary material. Relatedly, the present evaluation rests on a single train/test split, and the feature representation is limited to a compact, interpretable set of lexical, readability, and emotion features rather than contextual embeddings; repeated splits, bootstrapped confidence intervals, and transformer-based baselines would strengthen the robustness of these claims and are deferred to the supplement \cite{ibenye2026supplementary}. The datasets used here may also not fully reflect the diversity of real-world news or the evolving capabilities of generative models, motivating cross-domain and cross-model validation as a further direction. Purely text-based, surface-level detection may in any case face inherent limits against high-quality human-generated misinformation, suggesting that contextual, social, and network-based signals \cite{r5,r28} are an important complement to the linguistic features studied here.

\section{Conclusion}
Under an identical feature representation and modeling pipeline, AI-generated fake news is currently substantially more separable from real news than human-written fake news, across all six classifiers evaluated. This asymmetry suggests current detection systems are relatively well-equipped for today's AI-generated misinformation but less reliable against sophisticated human-crafted deception, motivating source-aware detection strategies. The relative difficulty of detecting human-written fake news also underscores a persistent and fundamentally challenging problem: human authors can strategically manipulate linguistic and contextual cues to mimic credible sources, making their outputs inherently closer to the distribution of real news. This suggests that purely text-based detection approaches, particularly those relying on surface-level linguistic features, may face inherent limitations in distinguishing high-quality human-generated misinformation from legitimate content, and that incorporating contextual, social, and network-based signals may be essential for closing this gap \cite{r5,r28, ghosh2025bot}.

Because the reported separability may partly reflect authorship artifacts rather than fakeness alone, we treat this as a controlled empirical baseline rather than a final characterization, and outline in the supplement a four-way experimental design, richer dataset and generation reporting, and additional baselines to isolate the effect of veracity from authorship. From a broader perspective, this study provides a controlled framework for comparing misinformation sources and emphasizes the importance of the underlying data-generating process: as generative models continue to improve, the observed gap may narrow, requiring detection methods that are more robust, adaptive, and capable of incorporating richer contextual and network-based information.

\paragraph*{\textbf{Acknowledgment}}
The authors thank the anonymous reviewers for comments that improved the clarity and rigor of this work.

\bibliography{ref}
\bibliographystyle{IEEEtran}

\end{document}